\documentclass[aps,pre,twocolumn,superscriptaddress,longbibliography]{revtex4-2}
\usepackage[utf8]{inputenc}
\usepackage{graphicx}
\usepackage{epstopdf}
\usepackage{bm}
\usepackage{amsmath}
\usepackage{verbatim}
\usepackage{amssymb}
\usepackage{listings}
\usepackage{color}
\definecolor{darkblue}{rgb}{0.1,0.1,0.8}
\usepackage{here}
\usepackage[colorlinks,linkcolor=darkblue,citecolor=darkblue,urlcolor=darkblue]{hyperref}
\renewcommand\vec[1]{\boldsymbol{#1}}

\newcommand\fig[1]{Fig.~\ref{#1}}
\newcommand\sect[1]{Sec.~\ref{#1}}

\renewcommand\vec[1]{\boldsymbol{#1}}

\newcommand\eq[1]{Eq.~(\ref{#1})}

\newcommand\average[1]{\left\langle #1 \right\rangle}
\newcommand\bigpar[1]{\left( #1 \right)}

\begin{document}

\title{Collective dynamics in a glass-former with Mari-Kurchan interactions}

\author{Yoshihiko Nishikawa}
\affiliation{Laboratoire Charles Coulomb (L2C), Universit\'e de Montpellier, CNRS, 34095 Montpellier, France}

\affiliation{Graduate School of Information Sciences, Tohoku University, Sendai 980-8579, Japan}

\author{Atsushi Ikeda}

\affiliation{Graduate School of Arts and Sciences, The University of Tokyo, Tokyo 153-8902, Japan}

\author{Ludovic Berthier}

\affiliation{Laboratoire Charles Coulomb (L2C), Universit\'e de Montpellier, CNRS, 34095 Montpellier, France}

\affiliation{Yusuf Hamied Department of Chemistry, University of Cambridge, Lensfield Road, Cambridge CB2 1EW, UK}

\date{\today}

\begin{abstract}
We numerically study the equilibrium relaxation dynamics of a two-dimensional Mari-Kurchan glass model. The tree-like structure of particle interactions forbids both non-trivial structural motifs and the emergence of a complex free-energy landscape leading to a thermodynamic glass transition, while the finite-dimensional nature of the model prevents the existence of a mode-coupling singularity. Nevertheless, the equilibrium relaxation dynamics is shown to be in excellent agreement with simulations performed in conventional glass-formers. Averaged time-correlation functions display a phenomenology typical of supercooled liquids, including the emergence of an excess signal in relaxation spectra at intermediate frequencies. We show that this evolution is accompanied by strong signatures of collective and heterogeneous dynamics which cannot be interpreted in terms of single particle hopping and emerge from dynamic facilitation. Our study demonstrates that an off-lattice interacting particle model with extremely simple structural correlations displays quantitatively realistic glassy dynamics. 
\end{abstract}

\maketitle

\section{Introduction}

Glasses are disordered solids formed by cooling a liquid across the glass transition~\cite{Debenedetti2001}. The resulting amorphous nature of glasses yields physical properties fundamentally different from the ones of crystals~\cite{doi:10.1063/PT.3.3052}. The dynamics of liquids approaching the glass transition is also different from the one of simple liquids, although their structures appear quite similar. The drastic increase of the structural relaxation time with a slight decrease of the temperature is accompanied by remarkable dynamic signatures~\cite{doi:10.1146/annurev.physchem.51.1.99,berthier2011dynamical,berthier2011dynamic}. Research on the glass transition aims at determining the microscopic origin of these unique dynamics in order to find the best theoretical framework to interpret them~\cite{Berthier2011}. 
 
Various theoretical approaches have attempted to interpret the glassy dynamics using thermodynamics, starting with the Adam-Gibbs theory~\cite{Adam1965}. The random first-order transition (RFOT) theory, which is constructed upon the mean-field theory for the glass transition~\cite{Kirkpatrick1987,Kirkpatrick1989,Kurchan2012,Charbonneau2014a,parisi2020theory} suggests similarly that the evolution of a complex free-energy landscape controls glassy dynamics. In the mean-field limit, a dynamic transition takes place when long-lived metastable states first appear, which bears similarity with the transition predicted by mode-coupling theory (MCT)~\cite{Kirkpatrick1987,gotze2009complex,PhysRevLett.116.015902}. In finite dimensions, infinitely long-lived metastable states cannot exist, and the MCT transition is avoided~\cite{Kirkpatrick1989,doi:10.1063/1.1796231}. Below this crossover, RFOT theory proposes that cooperative thermally activated events take place over a lengthscale of thermodynamic origin which grows at low temperatures and possibly diverges at the Kauzmann transition~\cite{Kirkpatrick1989,doi:10.1063/1.1796231}. In other structural approaches, real space structures~\cite{ROYALL20151} and geometric frustration~\cite{Kivelson1995,Tarjus2005} play preponderant roles in the description of glassy dynamics.

A different physical picture is provided by dynamical facilitation theory~\cite{Garrahan2002,Elmatad2009,Chandler2010}, which explains glassy dynamics from a purely kinetic perspective based on particle motion. In this view, the relaxation of supercooled liquids is controlled by a population of dynamical defects, defined as particles that are much more mobile than the bulk. In addition, mobile particles facilitate the displacement of their neighbors to create, at much later times, a mobility field that is correlated over a lengthscale that grows upon lowering the temperature~\cite{Butler1991,Garrahan2002,Whitelam2004}. Here, no static lengthscale is invoked to explain the correlated relaxation dynamics. This view is largely inspired by kinetically constrained glass models~\cite{Fredrickson1984,Jackle1991,doi:10.1080/0001873031000093582}. 

Despite their different underlying microscopic pictures, these viewpoints provide reasonable descriptions of a range of experimental and numerical data. It has repeatedly proven difficult to detect experimental signatures which can sharply favor one view against another. As a result, a complete and unified theoretical understanding of glassy dynamics is still lacking~\cite{tarjus2011overview,arceri2020glasses}.

Computer simulations are by definition well suited to test these theoretical concepts as the motion of all particles can be followed at all times. However, it is difficult to simulate systems at low enough temperatures to cleanly disentangle the role of local structures, mode-coupling crossover, thermodynamic fluctuations and dynamic facilitation in the simulated trajectories for finite-dimensional glass-formers.   

This problematic situation changed recently with the development of the swap Monte Carlo algorithm~\cite{Grigera2001,Berthier2016,Ninarello2017,Berthier_2019} which permits easy access to equilibrated configurations at essentially any of the temperatures that experiments can also explore. Whereas several novel features of ultrastable glasses have been revealed~\cite{doi:10.1073/pnas.1706860114,Ozawa2018,Berthier2019a,PhysRevE.102.042129,wang2019low,PhysRevLett.124.225901}, the remarkable speedup offered by swap Monte Carlo also provoked a debate regarding the relative role played by kinetic constraints and thermodynamics in glassy dynamics~\cite{Wyart2017,doi:10.1063/1.5009116,Berthier2019}. Swap Monte Carlo also opens a new window into the dynamics of deeply supercooled liquids, but again support for both thermodynamic~\cite{PhysRevLett.127.088002} and kinetic~\cite{guiselin2021microscopic} pictures were recently reported.

Another approach is to intentionally engineer artificial situations where only some of the ingredients assumed by theory survive or dominate. One can then ask whether and how dynamics is affected. If the dynamics is truly different from the one of conventional glass models, this suggests that the excluded factor plays an important role in realistic situations. There are many previous examples of this strategy where, for instance, the curvature of space~\cite{Sausset2008,Turci2017}, the number of space dimensions~\cite{Eaves2009,Berthier2020,adhikari2021spatial}, the number and geometry of freely-moving particles~\cite{biroli2008thermodynamic,Berthier2012,kob2012non,Kob2013}, or the range and strength of pair interactions~\cite{Berthier2009,Berthier2012,Berthier2009glass,ikeda2011gcm} have been varied in some arbitrary way to inform the physics.  

Here we follow this general strategy and analyse the dynamics of a two-dimensional Mari-Kurchan (MK) model~\cite{Mari2009,Mari2011} over a broad range of temperatures. The MK model is characterised by short-range pair interactions in a continuous space in finite dimension similar to conventional models, but interparticle distances are defined using an infinite-range random shift which suppresses many-body correlations~\cite{doi:10.1063/1.1724248,Mari2011}. This produces an artificial off-lattice particle model with short-range interactions in finite dimensions where: 1) the dynamic mode-coupling transition is strongly avoided~\cite{Charbonneau2014}; 2) the thermodynamics is very simple and the model cannot display a Kauzmann transition~\cite{Mari2011}; 3) there is no crystalline and locally-favored structures and geometric frustration should play a limited role; 4) equilibrium configurations of the model can easily be produced numerically at any temperature using a planting technique~\cite{Charbonneau2014}. Among the aforementioned physical ingredients, only dynamic facilitation can \textit{a priori} survive in the MK model. Of course, new relaxation channels with no finite dimensional analog could also emerge. It has in particular been argued that single particle hopping could become the relevant pathway for structural relaxation in the MK model~\cite{Charbonneau2014,Jin2015,biroli2021mean}. 

We find that the glassy dynamics of the MK model is extremely similar to results obtained in conventional models. It displays for instance a super-Arrhenius increase of the relaxation time, heterogeneous dynamics, and time correlation functions nearly indistinguishable from conventional glass-formers. We find in particular that particle mobility becomes increasingly correlated as the dynamics slows down, in a way that is incompatible with single particle hopping. We conclude that dynamic facilitation plays a central role in this model, and that dynamic facilitation alone can then lead to consistent physical behaviour.  

The rest of the paper is organized as follows. In \sect{sec:model}, we introduce the model and describe our simulation methods. We discuss time correlation functions, relaxation times, and relaxation spectra in \sect{sec:overlap}. We then discuss the heterogeneous dynamics at the single-particle level in \sect{sec:vanHove}. In \sect{sec:chi4}, we report the dynamical susceptibility and spatial correlation functions characterizing correlated particle dynamics. We summarize and discuss our results in \sect{sec:summary}.

\section{Model and methods}

\label{sec:model}

We study a two-dimensional Mari-Kurchan model with Hamiltonian~\cite{Mari2011} 
\begin{equation}
    H(\{\vec r_i\}) = \sum_{i<j} V\bigpar{|\vec r_i - \vec r_j - \vec A_{ij}|},
\label{eq:ham}
  \end{equation}
where $\vec r_i$ is the position of particle $i$, $\vec A$ is a real antisymmetric matrix with components drawn from a uniform distribution $\text{Unif}(0, L_\text{box})${, which we call the random shifts throughout the paper}. We denote by $L_\text{box}$ the linear dimension of the system and fix the volume fraction $\phi = N\pi (\sigma/2)^2 / L_\text{box}^2 = 2$, with $N$ the number of particles {and $\sigma$ the particle diameter}. We use a soft-sphere repulsive power law potential 
\begin{equation}
    V(r) / \varepsilon = 
    \bigpar{\frac{\sigma}{r}}^n + c_0 + c_2 \bigpar{\frac{r}{\sigma}}^2 + c_4 \bigpar{\frac{r}{\sigma}}^4, 
\end{equation}
if $r \leq r_\text{cutoff}$ and $V(r) = 0$ otherwise. Here, we set $n = 4$ and the constant values $c_i$ ($i=0,2,4$) are chosen so that $V(r) = V^\prime(r) = V^{\prime\prime}(r) = 0$ at the cutoff distance $r_\text{cutoff} = 2.5\sigma$. With these parameters, mean-field replica theory~\cite{doi:10.1063/1.5009116,nishikawa2021relaxation} yields the dynamical transition temperature $T_{d} \simeq 0.5940$. We show numerical data of the system with $N=1024$ particles in all figures, unless the system size is explicitly mentioned.

We simulate the underdamped Langevin dynamics at temperature $T$
\begin{equation}
    m\frac{d\vec v_i}{dt} = -\frac{\partial H}{\partial \vec r_i} - m\gamma \vec v_i + \sqrt{2m\gamma T}\vec \eta_i(t),
    \label{eq:langevin}
\end{equation}
with $\vec v_i$ the velocity of particle $i$, $m$ the particle mass, $\gamma$ the friction coefficient, and $\vec \eta_i$ the random force acting on particle $i$ having $\langle \vec \eta_i(t) \rangle = \vec 0$ and $\langle \eta_{i\mu}(0) \eta_{i\nu}(t)\rangle = \delta_{ij} \delta_{\mu\nu} \delta(t)$, where $\eta_{i\mu}(t)$ is the $\mu$-component of $\vec \eta_i(t)$ ($\mu = x, y$). {We measure time in the time unit of the dynamics $\tau_0 = \sqrt{m\sigma^2/\varepsilon}$, temperatures in the unit of $\varepsilon$ and distances in the unit of $\sigma$, thus they are dimensionless in the following}. We set $\gamma \tau_0 = 0.1$, which allows a stable numerical solution. In \eq{eq:langevin}, the average particle velocity scales as $\sim T^{1/2}$, and the microscopic time scale obviously slows down as $\sim T^{-1/2}$. We thus rescale the time $t$ by $T^{1/2}$ and plot time-dependent quantities as a function of {$T^{1/2}t$} when comparing different temperatures.

We can easily generate equilibrium configurations of the MK model at arbitrary temperature using the planting technique~\cite{Achlioptas2005,Achlioptas2008,Krzakala2009}. In practice this amounts to sampling the matrix $\vec A_{ij}$ for a given particle configuration $\{\vec r_i\}$ so that it becomes an equilibrium configuration of the Hamiltonian \eq{eq:ham}. The mismatch between the isotropic soft-sphere potential and the square box in which the particles reside complicates direct sampling of $\vec A_{ij}$ from the Boltzmann distribution. To deal with this complication, we use a Markov-chain Monte Carlo method with a simple Metropolis algorithm to sample $\vec A_{ij}$~\cite{nishikawa2021relaxation}. We perform $200$ Monte Carlo sweeps per pair of particles to ensure convergence to the Boltzmann distribution, and take the last configuration of $\vec A$ as the random shift for the chosen particle configuration.

The numerical method used to prepare equilibrium configurations is efficient and easy, so that preparing initial conditions remains extremely fast compared to the time needed to numerically study the low-temperature dynamics of the system. Our numerical strategy is thus to prepare a large number of independent initial conditions at each temperature, typically a hundred, from which we run the equilibrium dynamics. This allows us to analyse the equilibrium dynamics of the model down to arbitrarily low temperatures, including conditions where the equilibrium relaxation time is much longer than the simulated time. Conceptually, the planting technique advantageously replaces for the MK model the swap Monte Carlo technique~\cite{Berthier2016,Ninarello2017} used in finite dimensional models to analyse the low-temperature equilibrium dynamics~\cite{guiselin2021microscopic}.

A drawback of the MK model is the long-range nature of the matrix $\vec A_{ij}$. While this is key to endow the MK model with mean-field structure, this also penalises the numerical integration of the equations of motion \eq{eq:langevin} in two key places. First it slows down the creation of neighbor lists which requires the entries of the large matrix $\vec A_{ij}$ to be accessed. Second, storing the matrix itself may create memory problems for very large systems.    

Finally, the random shifts in the Hamiltonian (\ref{eq:ham}) imply that a snapshot of the system using the actual particle positions ${\vec r}_i$ would faithfully represent the positions of each particle, but would incorrectly represent interparticles distances which are not given by ${\vec r}_i -{\vec r}_j$ but instead by the shifted quantities ${\vec r}_i -{\vec r}_j -{\vec A}_{ij}$. It is therefore impossible to visualise all particle interactions simultaneously in real space. In particular, we will not be able to propose meaningful visual illustrations of the spatially heterogeneous dynamics of the MK model. 

\section{Correlation functions in time and Fourier domains}

\label{sec:overlap}

We first discuss the averaged dynamic behaviour of the MK model by following the evolution of correlation functions in both time and Fourier domains.

\subsection{Time correlation functions}

To study the glassy dynamics of the model, we measure the overlap between an initial configuration at $t=0$ and a configuration at time $t>0$ later
\begin{equation}
    \average{C(t)} = \average{\frac1N\sum_i C_i(t)},
\label{eq:overlap}
  \end{equation}
where $C_i(t) = \theta(|\vec r_i(t) - \vec r_i(0)| - \ell)$ {with $\theta(x)$ the Heaviside step function}. We also measure the self-intermediate scattering function
\begin{equation}
    \average{F_\text s(t)} = 
    \average{\frac1{dN} \sum_{i} \sum_\alpha \cos \left[ k(r_{i\alpha}(t) -r_{i\alpha}(0))\right] }
\label{eq:isf}
  \end{equation}
where $d=2$ is the spatial dimension, $k$ is a wavevector set to $k=\pi / (2 \ell)$, $r_{i\alpha}$ denotes the $\alpha$-component ($\alpha = x,y$) of the position of particle $i$. We denote the average over initial equilibrium configurations by the brackets $\average{\cdots}$. We shall take particles having displacements larger than {$1$} as `mobile' throughout this paper, which will be justified below in terms of the cage size observed in the mean-squared displacements and the van Hove distribution functions. Accordingly, we set $\ell = 1$ for the overlap function $C_i(t)$ in \eq{eq:overlap} and for the choice of wavevector in the self-intermediate scattering function in \eq{eq:isf}.

\begin{figure}[t]
    \centering
    \includegraphics[width=\linewidth]{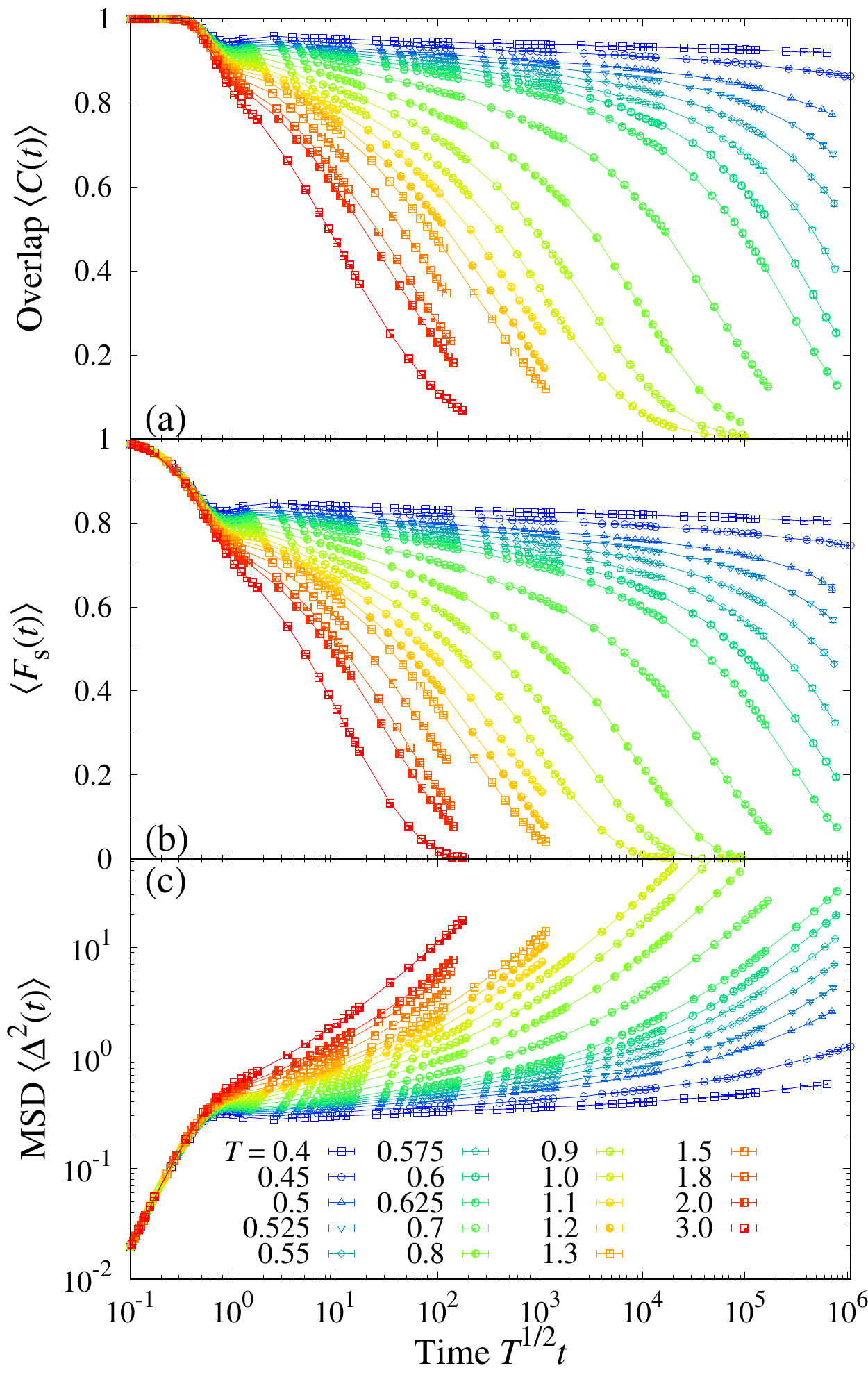}
    \caption{Time dependence of the overlap $\average{C(t)}$ (a), the self-intermediate scattering function $\average{F_\text s(t)}$ (b), and the MSD $\average{\Delta^2 (t)}$ (c) for various temperatures at thermal equilibrium. All functions develop a clear two-step decay below the onset temperature $T_{\rm onset}\approx 1.2$. A slow plateau dynamics can be observed at intermediate timescales in all cases.}
    \label{fig:overlap}
\end{figure}

We show in \fig{fig:overlap} the overlap $\average{C(t)}$ and the self-intermediate scattering function $\average{F_\text s(t)}$ of the MK model over a broad range of temperatures. At low temperature, both $\average{C(t)}$ and $\average{F_\text s(t)}$ show a clear two-step relaxation while they decay quickly at high temperatures. At $T < T_{d}$, the time correlation functions have a clear plateau, just as observed in finite-dimensional glass-formers. Closer inspection reveals that this is not strictly a constant plateau, as $\average{C(t)}$ and $\average{F_\text s(t)}$ decay very slowly at intermediate times between microscopic relaxation and structural relaxation. We observe the same slow decay in a larger system size $N=4096$, and conclude that this is not a finite size effect. This slow decay much before the structural relaxation indicates that a small fraction of particles are not well confined in cages and escape their initial position much before the majority of particles. This observation already suggests that the dynamics of the MK should be highly heterogeneous. A similar slow decay in the plateau region has recently been observed in finite-dimensional models~\cite{guiselin2021microscopic}, and associated with the emergence of excess wings in relaxation spectra.

We also measure the mean-squared displacement (MSD)
\begin{equation}
    \average{\Delta^2(t)} = \average{\frac1N\sum_i |\vec r_i(t) - \vec r_i(0)|^2}, 
\end{equation}
see Fig.~\ref{fig:overlap}(c). The overall evolution is again consistent with finite-dimensional systems with a diffusive regime at long times that becomes very slow at low temperatures and the development of a plateau at intermediate times. In agreement with the overlap and self-intermediate scattering function, the MSD also shows a slow increase with time in the plateau region. Again, this slow time dependence is observed in the larger system and should exist in the thermodynamic limit. The plateau height of the MSD at $T \simeq T_{d}$ is $\average{\Delta^2} \simeq 0.4$ corresponding to a cage size $\simeq 0.63$, justifying our choice for the value {$\ell = 1$} above for the definition of the overlap and the self-intermediate scattering function.

\subsection{Relaxation times and characteristic temperatures}

\begin{figure}[t]
    \includegraphics[width=\linewidth]{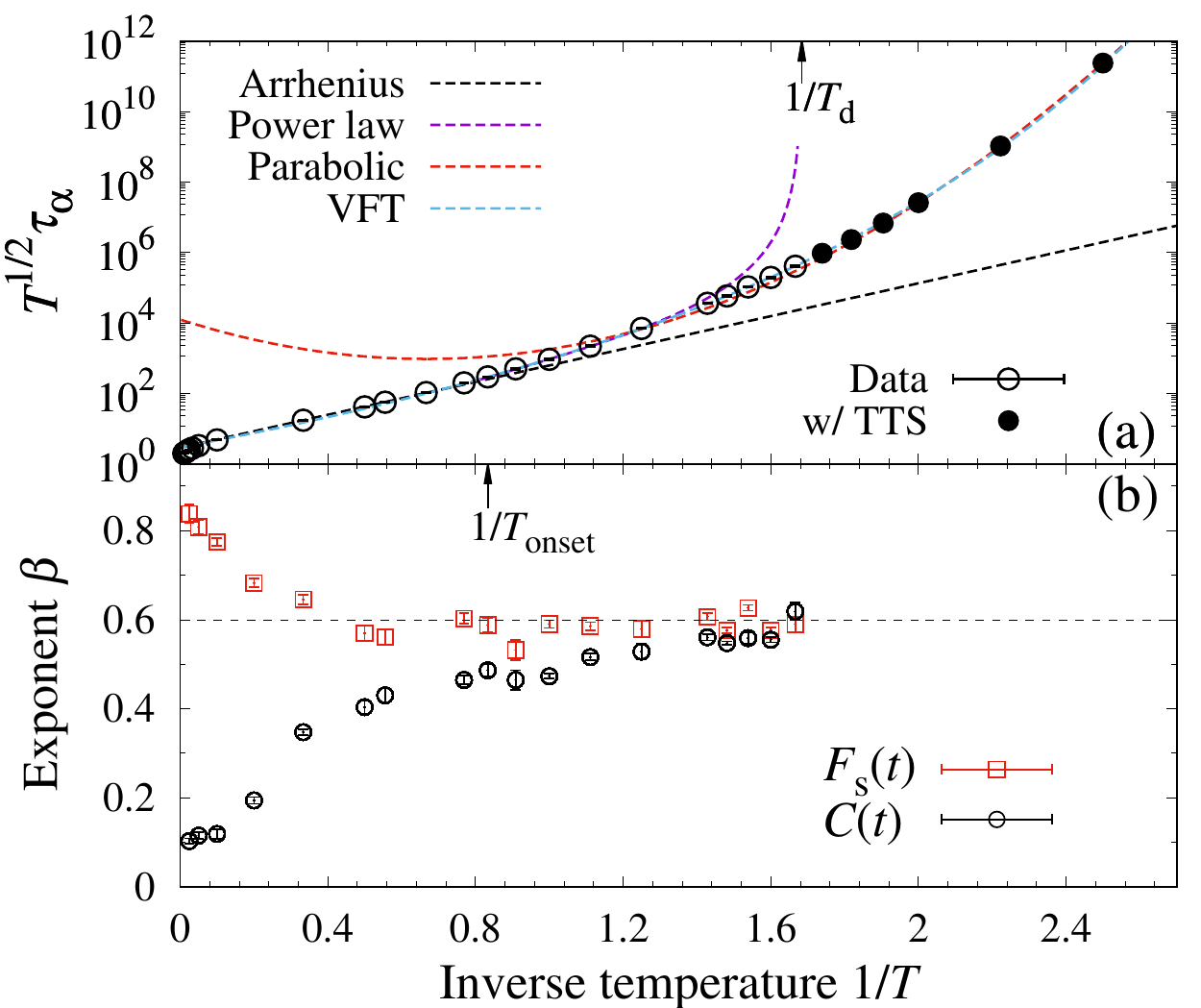}
    \caption{(a) Temperature evolution of the normalised relaxation time, together with various fits. High temperature Arrhenius regime, $\sim \exp(\Delta E / T)$ with $\Delta E = 5.3$; power-law divergence $\sim (T - T_{d})^{-\gamma}$ with $T_{d} = 0.5940$ and $\gamma = 2.97$. Low temperatures are equally well fitted with the parabolic law $\sim \exp((T_\text{o}/T - 1)^2)$ with $T_\text{o} = 1.49$, and the VFT law $\sim \exp(A/(T - T_0))$ with $T_0 = 0.22$. Using $\average{C(t)}$ at temperature $T < 0.6$, we estimate the relaxation time using time-temperature superposition, see black circles.
      (b) Temperature evolution of the stretching exponents $\beta$ for $\average{C(t)}$ and $\average{F_{\text s}(t)}$, converging to $\beta \approx 0.6$ at low $T$ (dashed line).}
    \label{fig:tau}
\end{figure}

We now extract the $\alpha$-relaxation time $\tau_\alpha$ which we define as the time where $\average{C(t)}$ reaches $1/e$. For the lowest temperatures at which $\average{C(t)}$ clearly decays from its plateau but does not reach $1/e$ in our simulation time window, we assume that time-temperature superposition (TTS) holds to estimate $\tau_\alpha$~\cite{doi:10.1063/5.0015227}. We report the results of this analysis in \fig{fig:tau}(a).

At low temperatures where the two-step relaxation becomes pronounced, we observe that $\tau_\alpha$ increases rapidly and shows a super-Arrhenius dependence on the temperature. Fitting the high temperature regime to an Arrhenius law provides an estimate of the onset of glassy dynamics for this system near $T_{\rm onset} \approx 1.2$, below which the Arrhenius description does not hold.  

We recall that the mean-field replica theory yields the dynamical transition temperature $T_{d} \simeq 0.5940$. The mean-field theory of the glass transition predicts that the relaxation time shows a critical divergence at $T_{d}$. However, around this temperature, the relaxation time is almost independent of the system size and no critical divergence is observed, see Fig.~\ref{fig:tau}(a). This indicates the MCT transition is strongly avoided for this system~\cite{Charbonneau2014}.

On the other hand, the relaxation time is well fitted by the parabolic law~\cite{Elmatad2009,Chandler2010} with an onset temperature $T_\text{o} \simeq 1.49$, which appears consistent with a dynamic facilitation description. Interestingly, even though the MK model has no Kauzmann transition at any finite temperature, the Vogel-Fulcher-Tammann (VFT) law~\cite{Berthier2011,Cavagna2009} with a critical temperature $T_0 \simeq 0.22$ fits the relaxation time just as well as the parabolic law, see \fig{fig:tau}(a). This implies that, similarly to finite-dimensional systems, it is impossible to distinguish between these two functional forms reliably using solely the temperature dependence of the relaxation time~\cite{hecksher2008little}.

We also characterize the shape of the $\alpha$-relaxation observed in the functions $\average{C(t)}$ and $\average{F_s(t)}$ by fitting their long-time decay to a stretched exponential form $\sim \exp(-(t / \tau)^\beta)$. As shown in \fig{fig:tau}(b), the measured exponent $\beta(T)$ for these two correlations converge to a similar low-temperature value $\beta \simeq 0.6$ when the temperature decreases below $T_\text{onset}$. A common interpretation for the stretching exponent $\beta<1$ is that the underlying dynamics is highly heterogeneous~\cite{doi:10.1146/annurev.physchem.51.1.99}. Note that in the high-temperature limit where the dynamics is expected to be homogeneous and diffusive, the self-intermediate function becomes a simple exponential with $\beta \approx 1$, but the overlap $\average{C(t)}$ is never a simple exponential. This is a direct consequence of its mathematical definition involving the Heaviside function $\theta(x)$ which leads to a power law decay for pure Fickian dynamics, explaining the low $\beta$ values extracted at high temperatures for the overlap. Both functions become however equivalent when they are controlled by a broad underlying distribution of relaxation times. Also, the convergence of the stretching exponent $\beta$ at low temperatures to a nearly constant value justifies our use of time-temperature superposition to extrapolate the relaxation time at lower temperatures, as shown in Fig.~\ref{fig:tau}(a). Similar findings were recently reported for several glass-forming models~\cite{guiselin2021microscopic,PhysRevLett.127.088002,doi:10.1063/5.0015227}. 

\subsection{Spectra in the frequency domain}

To analyze more precisely the plateau dynamics detected in the time correlation functions, it is convenient to perform this analysis in the frequency domain because the large constant value of the plateau automatically disappears and only the small, interesting time dependence contributes. 

We calculate the analog of the imaginary part of the dielectric loss, i.e. the relaxation spectrum~\cite{Williams1972,Jambeck2015},
\begin{equation}
    \chi^{\prime\prime}(\omega) =
  - \int d\log \tau G(\log \tau)\frac{\omega \tau}{1 + (\omega \tau)^2},
  \label{eq:spectra1}
\end{equation}
where $G(\log \tau)$ is a timescale distribution satisfying
\begin{equation}
    \average{C(t)} = \int d\log\tau G(\log \tau) \exp\bigpar{-\frac{t}{\tau}}.
\end{equation}
Unfortunately, estimating $G(\log \tau)$ directly from $\average{C(t)}$ requires an inverse Laplace transform, which is notoriously difficult and numerically unstable. In the same vein, direct Fourier transform of $\langle C(t)\rangle$ is plagued by important statistical noise~\cite{guiselin2021microscopic}. Following earlier work~\cite{Berthier2005,guiselin2021microscopic}, we make the following assumption
\begin{equation}
    G(\log \tau) \simeq -\frac{d \average{C(\tau)}}{d\log\tau}
\end{equation}
and calculate 
\begin{equation}
    \chi^{\prime\prime}(\omega) = \int d\log\tau \frac{d\average{C(\tau)}}{d\log\tau}\frac{\omega \tau}{1 + (\omega \tau)^2},
    \label{eq:app_spectra}
\end{equation}
which is an excellent approximation to the desired relaxation spectrum. We numerically checked the approximation works quite well by comparing \eq{eq:app_spectra} and \eq{eq:spectra1} at small $\omega$. In addition, since we only collect data in a finite window of maximum duration $t_\text{sim}$, it is obvious that the precision of \eq{eq:app_spectra} is limited to $\omega \gg t_\text{sim}^{-1}$. At low temperatures $T=0.575$, $0.6$, and $0.65$ where $\average{C(t)}$ reached $1/e$ within the simulation time and the exponent $\beta$ is well approximated as $\beta=0.6$, we concatenate $\average{C(t)}$ and the fitted stretched-exponential decay to obtain $\chi^{\prime\prime}$ down to $\omega = 10^{-5}$. For the other temperatures, we show the numerical data over a range limited to $\omega \geq 10^2/ t_\text{sim}$.

\begin{figure}[t]
    \centering
    \includegraphics[width=\linewidth]{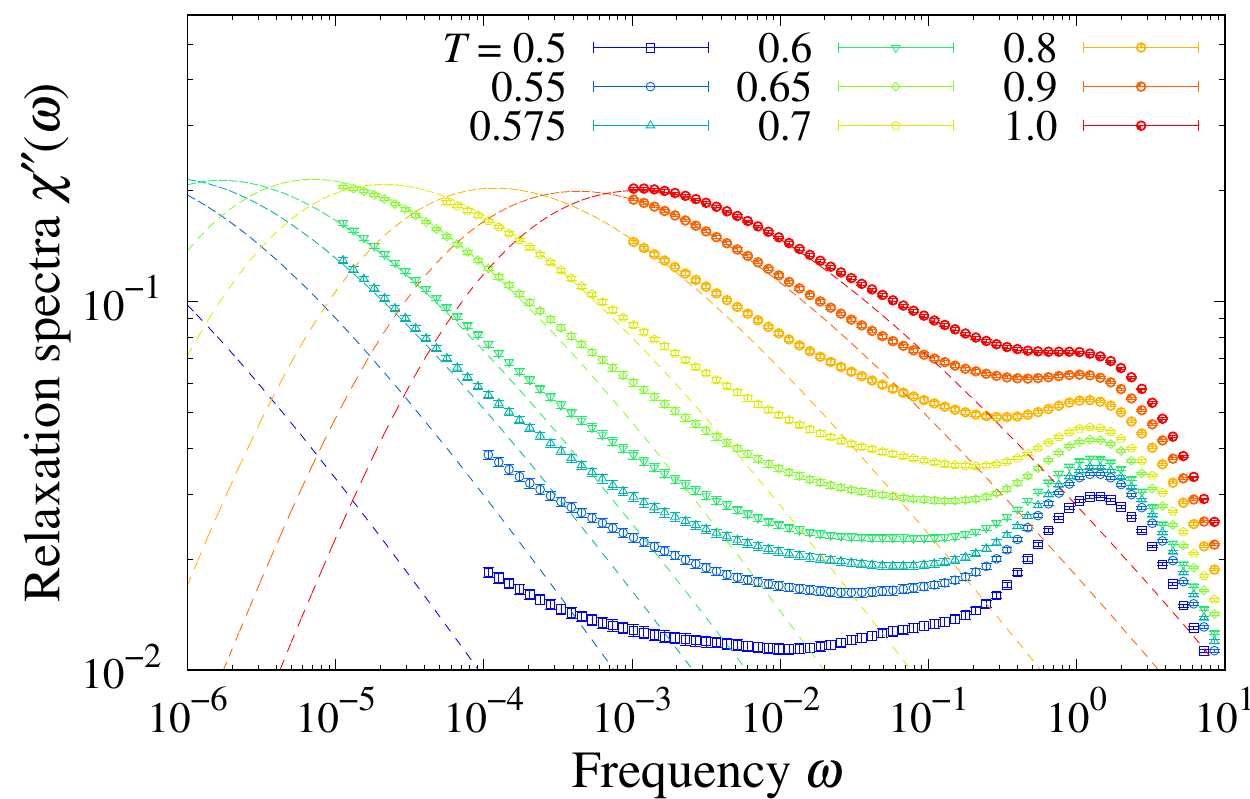}
    \caption{Relaxation spectra $\chi^{\prime\prime}(\omega)$ obtained from the overlap $\average{C(t)}$ and \eq{eq:app_spectra}. The peaks at $\omega \simeq 1$ and lower $\omega \simeq \tau_\alpha^{-1}(T)$ correspond to microscopic and structural relaxations, respectively. Dashed lines are obtained using stretched exponential fits to the overlap $\average{C(t)}$ in \eq{eq:app_spectra}. A clear excess signal appears at intermediate frequencies at low temperatures.}
    \label{fig:spectra}
\end{figure}

In \fig{fig:spectra} we show $\chi^{\prime\prime}(\omega)$ at several temperatures. As a reference, we also plot $\chi^{\prime\prime}(\omega)$ at much lower $\omega$ by applying \eq{eq:app_spectra} to the stretched exponential fits of $\average{C(t)}$ used to estimate the relaxation times at very low temperatures. The two-step decay of the time correlation functions gives rise to relaxation spectra in the frequency domain characterised by two well-separated peaks. The first peak at large frequency $\omega \simeq 1$ corresponds to the fast relaxation at microscopic times. The second peak shifts to lower frequencies at lower temperatures and corresponds to the structural $\alpha$-relaxation occurring near $\omega \simeq \tau_\alpha^{-1}(T)$~\cite{Pardo2007}.

In the frequency domain intermediate between these two peaks, the spectra at low temperatures are significantly larger than the linear superposition of these two processes. In particular, when comparing the measured spectra with the high-frequency extrapolation of the structural relaxation peak, we observe that all measured spectra show a signal which is significantly in `excess' of the main peak, see \fig{fig:spectra}. This intermediate frequency regime corresponds, in the time domain, to the plateau regime where $\average{C(t)}$ decays very slowly on timescales that are much shorter than the structural relaxation fitted with a stretched exponential decay. These measurements confirm that in the plateau regime at times $t \ll \tau_\alpha$, there is considerably more relaxation dynamics than expected from the short-time expansion of the streched exponential fit to the $\alpha$-relaxation. 

Although the signal that we measure at intermediate frequencies does not have a simple power-law dependence on $\omega$ at low temperatures, this excess signal is qualitatively similar to the excess wing observed in various experiments in molecular liquids~\cite{nagel_scaling,nagel_scaling2,leheny1997high,Schneider2000,Blochowicz2003,Blochowicz2006} and very recently detected numerically in glass-formers in two and three dimensions~\cite{guiselin2021microscopic}. From Fig.~\ref{fig:spectra} we observe that the excess signal could be fit with a very small power law that would seem to decrease with decreasing temperature, as observed in experiments~\cite{nagel_scaling,nagel_scaling2}. This is maybe consistent with the power law distribution of escape times measured numerically in Ref.~\cite{Charbonneau2014}. For the lowest temperature, the time evolution on the plateau of $\langle C(t) \rangle$ becomes nearly logarithmic, which would correspond to a nearly flat spectrum, as indeed observed in Fig.~\ref{fig:spectra}. One would need to simulate lower temperatures for longer times to fully assess the frequency dependence and temperature evolution of the excess signal in the relaxation spectra of the MK model. 

\section{Dynamic heterogeneity: Single particle dynamics}

\label{sec:vanHove}

We now move to the analysis of the heterogeneous nature of the dynamics, starting at the single particle level. To this end, we measure the probability distribution function of single particle displacements, which is the self-part of the van Hove distribution function~\cite{PhysRev.136.A405},
\begin{equation}
    P_t(\Delta x) = \average{\frac{1}{dN} \sum_{i}\sum_{\alpha} \delta \left[ \Delta x - \bigpar{r_{i\alpha}(t) - r_{i\alpha}(0)} \right] } .
\end{equation} 
At very high temperatures, where the system behaves as a simple fluid with diffusive behaviour, we expect $P_t(\Delta x)$ to be Gaussian with a variance increasing linearly with time. In glass-formers, deviations from this Gaussian behaviour typically indicate that the system is dynamically heterogeneous~\cite{PhysRevA.38.3758,doi:10.1126/science.287.5451.290,doi:10.1126/science.287.5453.627}.  

\begin{figure}[t]
    \centering
    \includegraphics[width=\linewidth]{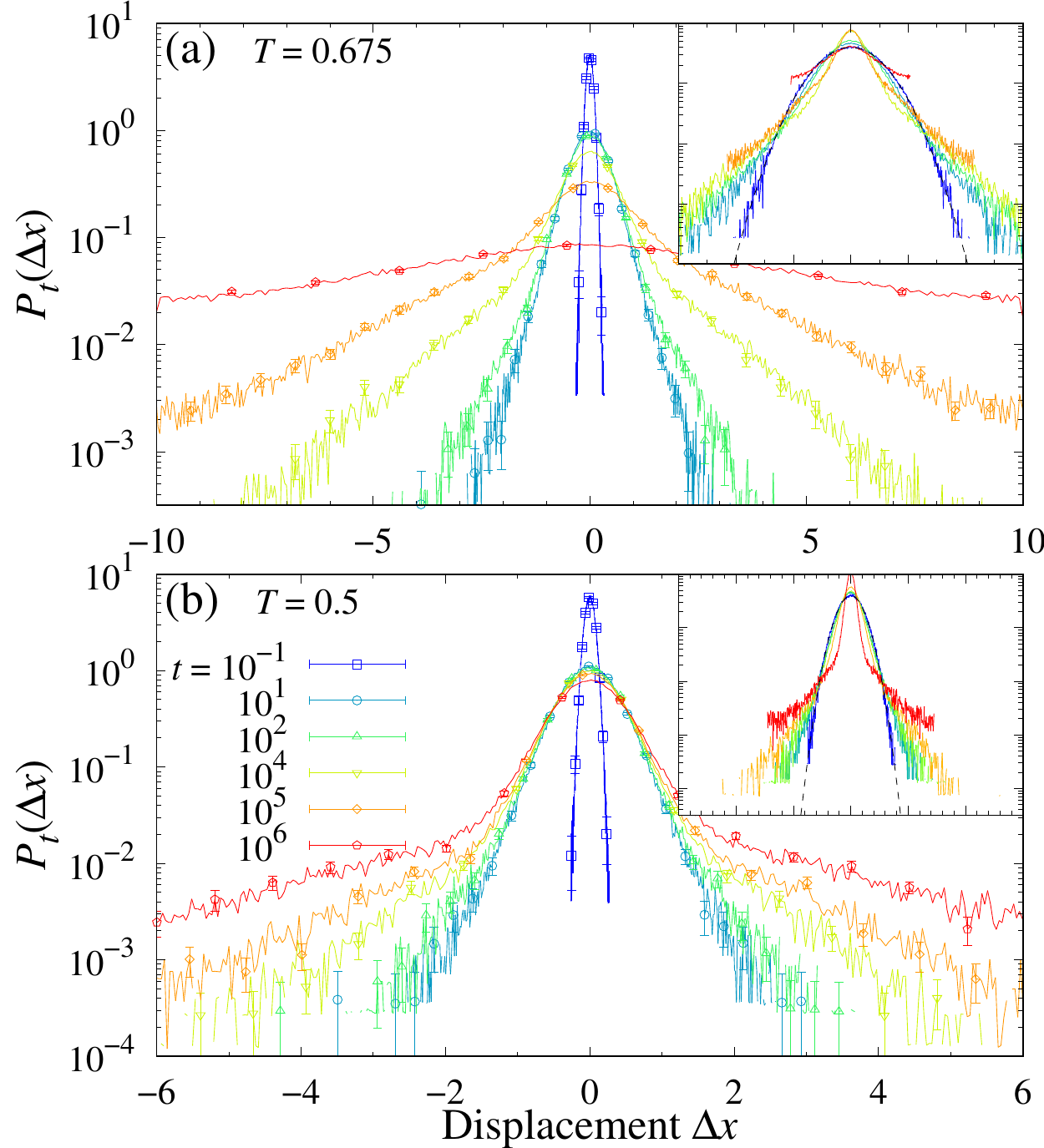}
    \caption{The van Hove distribution function $P_t(\Delta x)$ at (a) $T = 0.675$ and (b) $T=0.5$. Insets show the distributions scaled using the measured variance with the dashed line indicating the normal distribution. The structural relaxation time is  $\tau_\alpha \simeq 2.5 \times 10^5$ at $T = 0.675$ and $\simeq 5\times 10^{7}$ at $T=0.5$.}
    \label{fig:vanHove}
\end{figure}

We show the van Hove function $P_t(\Delta x)$ in \fig{fig:vanHove} at temperatures $T = 0.675$ and $0.5$. Measuring the variance $\lambda^2(t)$ of these distributions, we also replot $\lambda P_t$ as a function of $\lambda \Delta x$ in the insets of \fig{fig:vanHove}. These rescaled distributions become the normal distribution whenever $P_t(\Delta x)$ is Gaussian. This is the case at all timescales at high temperatures, $T \gg 1$, as expected.

At low temperatures instead, the distributions exhibit broad tails except at very small and very large times. Qualitatively, the van Hove distributions develop nearly exponential tails already in the plateau region, see \fig{fig:vanHove}. Such exponential tails decay much more slowly than the Gaussian prediction for Fickian dynamics, which reveals the existence of a small concentration of particles that have moved much farther than the majority after a given time. This non-Gaussian shape and exponentially decaying tails in the van Hove function have been universally observed in finite-dimensional glass-forming liquids~\cite{Chaudhuri2007}.

For the lowest temperature, $T=0.5$, the largest time $t=10^6$ is much shorter than the estimated relaxation time, $\tau_\alpha \sim 5 \times 10^7$ and most of these distributions thus correspond to the very extended plateau regime seen at this temperature. In this plateau regime, the distribution is almost invariant in its nearly Gaussian core, but the tails become broader and more populated with increasing time. The population of faster than average particles thus increases with time and moves, on average, over larger distances. This population of very fast particles is therefore responsible for the slow time decay of $\average{C(t)}$ in the plateau time region and of the excess relaxation signal detected in the relaxation spectra. This observation is again consistent with the microscopic interpretation of excess wings in finite dimensional glass-formers~\cite{guiselin2021microscopic,doi:10.1063/5.0060408}. 

\section{Dynamic heterogeneity: Collective dynamics}

\label{sec:chi4}

\subsection{Four-point dynamic susceptibility}

The analysis of the van Hove distributions in \sect{sec:vanHove} shows the coexistence of fast and slow moving particles at any given time. As is well known, however, the distribution of single particle displacements contains no information about possible correlations between particle displacements. Here, we analyse observables that can reveal this information and allow us to discuss the collective nature of the relaxation dynamics in the MK model. 

Using multiple independent trajectories, we can calculate the four-point dynamical susceptibility defined by the variance of the spontaneous fluctuations of the dynamic overlap~\cite{DONATI2002215,doi:10.1063/1.1605094,Toninelli2005}: 
\begin{equation}
    \chi_4(t) = N \frac{\average{C^2(t)} - \average{C(t)}^2}
    {\average{C(t)}(1 - \average{C(t)})}.
    \label{eq:chi4}
\end{equation}
In \eq{eq:chi4}, $\chi_4(t)$ is normalized such that $\chi_4(t)=1$ when the dynamics of all pairs of particles is completely uncorrelated.

\begin{figure}
    \centering
    \includegraphics[width=\linewidth]{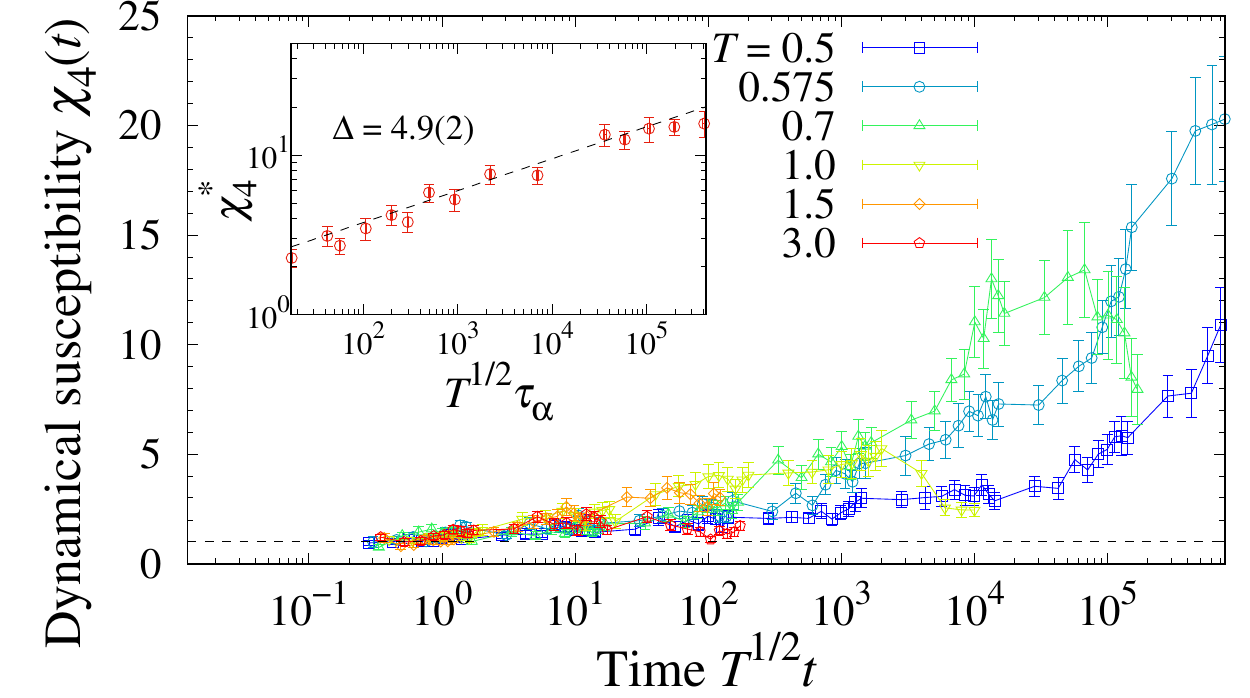}
    \caption{Dynamical susceptibility $\chi_4(t)$ at several temperatures. The dashed line is $\chi_4(t) = 1$, the trivial value observed for uncorrelated dynamics. Inset: maximal value $\chi^*_4$ as a function of rescaled structural relaxation time $T^{1/2} \tau_\alpha$, with dashed line  indicating the power law $\chi_4^* \sim ( T^{1/2} \tau_\alpha)^{1/\Delta}$ fitted to the numerical data with $\Delta = 4.9(2)$.}
    \label{fig:chi4}
\end{figure}

In \fig{fig:chi4} we show the dynamical susceptibility $\chi_4(t)$ at several temperatures. At very short time $T^{1/2} t \simeq 1$, $\chi_4(t) \simeq 1$, consistent with uncorrelated microscopic motion at microscopic times. On the time scale where time correlation functions have a plateau, $\chi_4(t)$ grows monotonically with time, possibly as a power law with a small exponent near 0.2. Therefore the increasing number of mobile particles at intermediate times revealed by the van Hove distribution cannot be interpreted as the independent hopping of particles out of their respective cages. Instead, the dynamics of mobile particles become increasingly correlated with time, and particle hopping is actually a correlated, many-body process.   

At times $t \simeq \tau_\alpha(T)$, $\chi_4(t)$ reaches a maximum value $\chi_4^*(T)$ which is significantly larger than $1$, before decreasing towards unity at much longer time scales $t \gg \tau_\alpha$. This time dependence is in agreement with the typical behaviour observed in finite-dimensional glass-forming liquids~\cite{Toninelli2005}. This similarity implies that the dynamics of the MK model is just as collective as the one of {conventional glass models}. Remarkably, even below the avoided dynamical transition temperature $T_{d}$, $\chi_4(t)$ grows with time and becomes much larger than $1$, contrary to the proposed picture of a relaxation process governed by the independent hopping of individual particles~\cite{Charbonneau2014}.

In the inset of \fig{fig:chi4}, we show the parametric evolution of $\chi_4^*(T)$ with the rescaled relaxation time $T^{1/2} \tau_\alpha(T)$ together with an algebraic dependence $\chi_4^* \sim \bigpar{T^{1/2} \tau_\alpha}^{1/\Delta}$ with $\Delta = 4.9(2)$. The exponent $\Delta$ is somewhat larger than that reported for finite-dimensional models~\cite{Whitelam2004,Toninelli2005,Brambilla2009,Berthier2011,Berthier2012}, {$\Delta \approx 2.5$,} but close to that of the hard-sphere MK model \cite{Mari2011}, {$\Delta \approx 4.5$}. This may be due to the fact that the mode-coupling crossover, which would yield the prediction $\chi_4^* \sim \tau_\alpha^{1/\gamma}$, is very strongly suppressed in the present MK model. As a result, the growth of $\chi_4^*$ may be solely due to collective activated events for which much slower growth, possibly logarithmic, is expected on general grounds~\cite{Toninelli2005} and indeed found in several cases~\cite{Berthier2005,Dalle-Ferrier2007}. In any case, the present data unambiguously reveals that the low-temperature dynamics of the MK model is actually governed by strong many-body correlations which do not stem from (avoided) mode-coupling criticality. 

\subsection{Spatial correlation functions}

The dynamic susceptibility $\chi_4(t)$ can also be represented as the integral of the four-point correlation function $G_4(r, t)$ over space:
\begin{equation}
    \chi_4(t) = \frac{\int dr G_4(r, t)}{G_4(0, t)},
\end{equation}
where 
\begin{equation}
    G_4(r, t) = \frac1V\average{\sum_{i, j} 
    \delta C_i(t)
    \delta C_j(t)
    \delta(r - r^{\vec A}_{ij}(0))},
    \label{eq:G4}
\end{equation}
with $\delta C_i(t) = C_i(t) - \average{C(t)}$ and $r^{\vec A}_{ij}(t) = |\vec r_i(t)- \vec r_j(t) - \vec A_{ij}|$. Thus, a large value of $\chi_4(t)$ implies the existence of large correlations between local fluctuations of the dynamics, $\langle \delta C_i(t) \delta C_j(t) \rangle$.  Recall however that the meaning of the spatial variable $r$ in \eq{eq:G4} does not have the ordinary interpretation as in conventional models due to the long-range random shifts in the definition of the Hamiltonian \eq{eq:ham}.

\begin{figure}
    \centering
    \includegraphics[width=\linewidth]{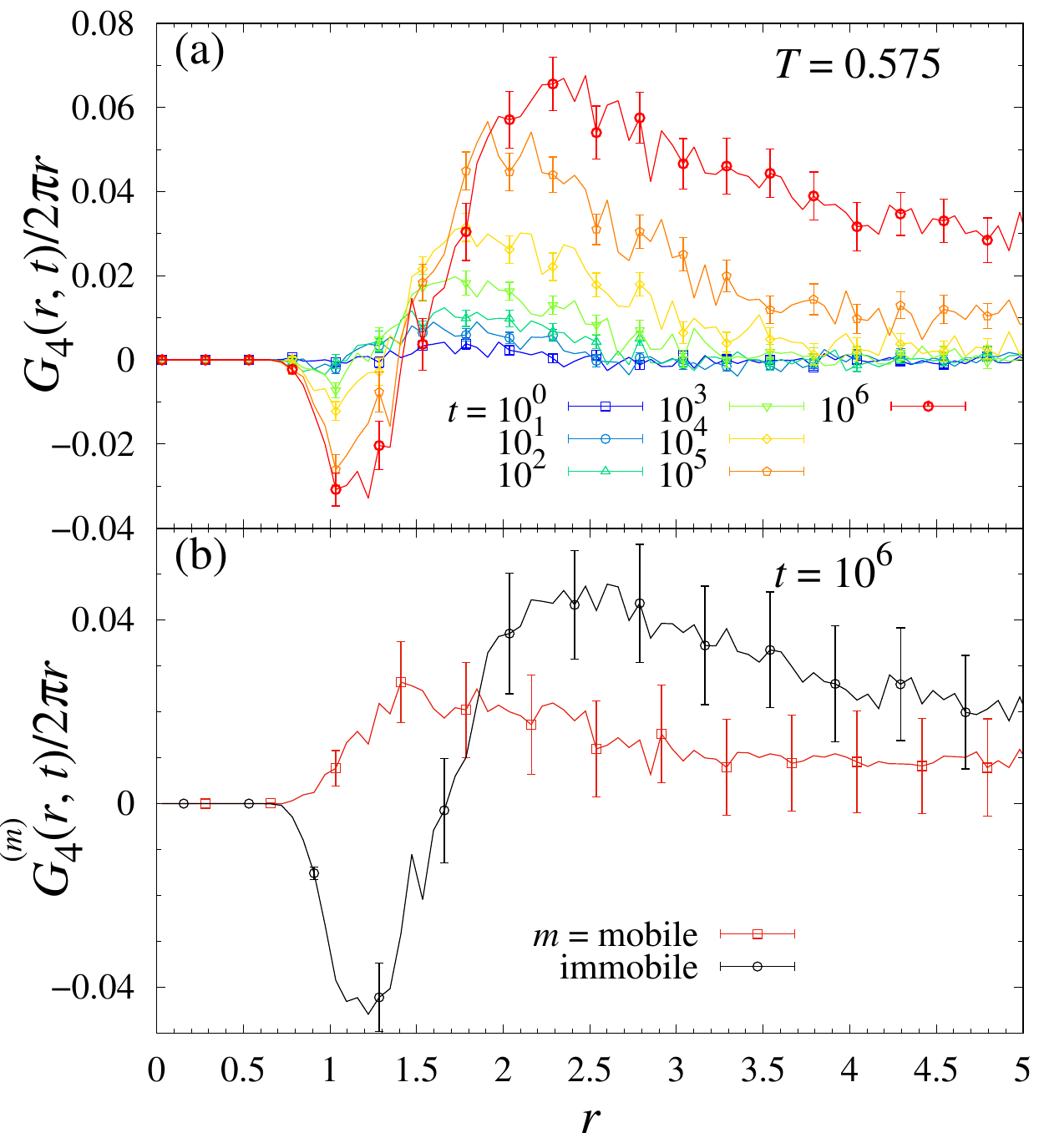}
    \caption{Four-point correlation function (a) $G_4(r, t) / 2\pi r$ for different times and (b) $G^{(m)}_4(r, t) / 2\pi r$ at $t = 10^6$, around which $\chi_4(t)$ is maximum. The temperature is $T = 0.575$ in both panels. }
    \label{fig:G4}
\end{figure}

Even if they are difficult to visualise, correlations between the dynamics of individual particles are anyway directly revealed in the four-point correlation $G_4(r, t)$. We show $G_4(r, t) / 2\pi r$ in \fig{fig:G4} at $T=0.575$. At shown in \fig{fig:G4}(a), $G_4(r, t)$ develops a clear peak at $r \simeq {2}$ which changes with time and temperature. At the temperature shown in \fig{fig:G4}(a), the relaxation time is $\tau_\alpha \simeq 10^6$ which is also roughly the time when $\chi_4(t)$ becomes maximum. The time dependence of $G_4(r, t)$ is compatible with this evolution.

However, $G_4(r, t)$ displays two features that are not seen in finite dimensional models. First, $G_4(r, t)$ has a negative dip at short distance $r \lesssim {1}$ at any time $t$. This may naively suggest that particles in very close contact have anticorrelated mobility, which appears counterintuitive at first sight. Second, $G_4(r, t)$ does not decay at large distances. We discuss these two points in the following. 

To understand the physical origin of the anticorrelation and the negative dip at small $r$, we decompose $G_4(r,t)$ into two independent parts, that is, mobile and immobile contributions to $G_4(r, t)$. Let us define
\begin{equation}
    \label{eq:mobileG4}
    G_4^{(m)}(r, t) = \frac1V\average{\sum_{i, j} 
    \alpha_i(t)
    \delta C_i(t)
    \delta C_j(t)
    \delta(r - r^{\vec A}_{ij}(t))},
\end{equation}
where $\alpha_i(t) = C_i(t)$ if $m = \text{mobile}$ and $\alpha_i = 1 - C_i(t)$ if $m = \text{immobile}$. Then by definition, we have $G_4(r, t) = G_4^\text{(mobile)}(r, t) + G_4^\text{(immobile)}(r, t)$. {We remind that particles with displacement larger than $1$ are defined as mobile and the rest as immobile.} We show the result of this decomposition in \fig{fig:G4}(b) at $T = 0.575$. Clearly, at $r \simeq {1}$, the immobile contribution is negative while the mobile one is positive, meaning that the anticorrelation observed for the total function comes solely from immobile particles. 

\begin{figure}[t]
    \centering
    \includegraphics[width=\linewidth]{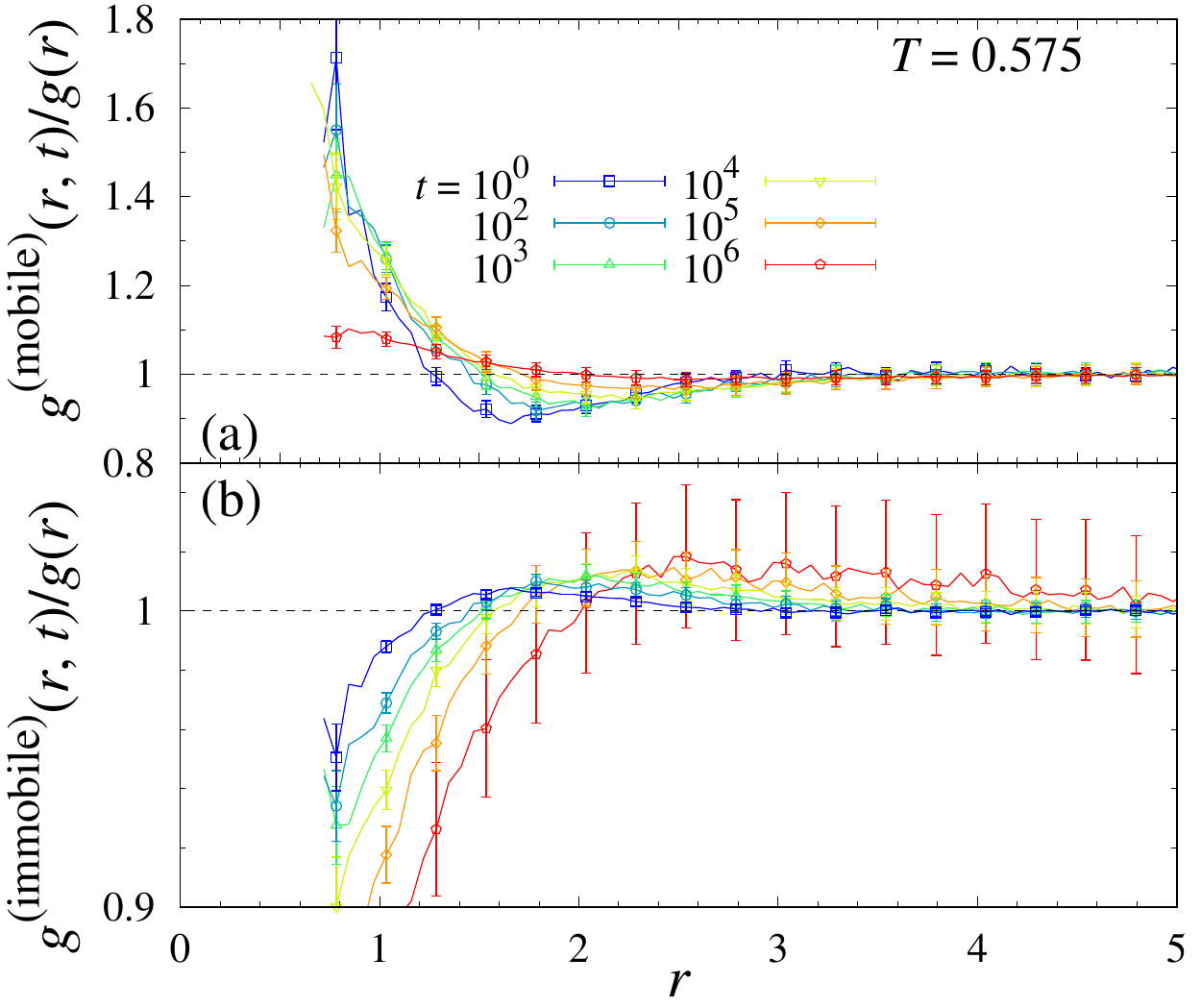}
    \caption{Radial distribution functions for (a) mobile and (b) immobile particles at $T = 0.575$ and different times. Mobile particles have more neighbors at small distances $r\simeq {1}$ and less at $r\simeq {2}$, while immobile particles have less at $r\simeq {1}$ and more at $r\simeq {2}$. This indicates a clear correlation between structure and dynamics.}
        \label{fig:m_im_gr}
\end{figure}

The clear difference between mobile and immobile contributions to $G_4(r,t)$ suggests that particle mobility originates, on average, from the local structure formed by surrounding particles. To test this idea, we measure the static radial distribution function separately for mobile and immobile particles:
\begin{equation}
g^{(m)}(r,t) = 
\frac{\average{\sum \alpha_i(t)\delta(r - r^{\vec A}_{ij}(t))}}{V\rho\average{\rho^{\text{(m)}}(t)}},
\end{equation}
where $\rho^{\text{(m)}}(t)$ is the number density of mobile (immobile) particles if $m = \text{mobile}$ ($\text{immobile}$). This radial distribution function characterizes the average surrounding structure of mobile or immobile particles. We normalise $g^{(m)}(r,t)$ by the bulk pair correlation function $g(r)$,
\begin{equation}
    \label{eq:gofr}
    g(r) = \average{\frac{V}{N^2}\sum_i \delta\bigpar{|\vec r_i - \vec r_j - \vec A_{ij}|}},
\end{equation}
so that differences between the local structure and the average one are directly revealed by a ratio that differs from unity. The results are shown in \fig{fig:m_im_gr}. We can see that mobile particles have more neighbors on average at short distances $r\lesssim {1}$, but less at large distances, $r \simeq {2}$. By contrast, immobile particles have less neighbors at short distances $r \lesssim {1}$, and slightly more at $r \simeq {2}$. We finally understand that the negative dip in the immobile contribution to $G_4(r,t)$ thus comes from the fact that immobile particles have less neighbors at short distances, and indeed does not point to the existence of unphysical mobility anticorrelations. 

Let us offer some physical interpretation of these findings. In supercooled liquids, particles are confined within a cage comprised of neighboring particles, and a particle becomes mobile by escaping this cage. In this view, the radial distributions for mobile particles indicating that they have more neighbors at short distance, and are thus more confined than average appears counterintuitive. We believe that the reason is that particles interacting with their neighbors at short distances are then subject to large forces, which potentially leads to a stronger force imbalance which can induce rapid displacements. In addition, \fig{fig:m_im_gr}(a) shows that mobile particles  have less particles at $r \simeq {2}$. Thus, once they escape their first shell of neighbors $r\lesssim {1}$, they are not efficiently pushed back to their original position by particles further away. The opposite effects apply to the immobile particles, see \fig{fig:m_im_gr}(b).

Regarding the absence of decay at large $r$, we recall that the distance $r$ used in the definition of $G_4(r,t)$ is meaningful for neighboring particles which effectively interact at short shifted distances $r_{ij}^{\vec A}$. However, the physical meaning of large shifted distances is lost. In particular, it is not possible to extract a dynamic correlation length from the four-point function $G_4(r,t)$. Still, the large value in $G_4(r,t)$ at large distances is consistent with the large value of $\chi_4(t)$, and it proves that dynamics is collective and involves multi-body correlations. This implies that the two-particle correlations detected by $G_4(r,t)$ at short distances must propagate in space and time to form large correlated clusters at $t \simeq \tau_\alpha$, even though we cannot visualise them in real space. 

To discuss the correlations to second and third neighbors and extract a  dynamic correlation length, a more advanced analysis considering the tree-like structure of interactions would be needed, which we leave for future work.

\section{Summary and Discussion}

\label{sec:summary}

To summarize, we studied the equilibrium dynamics of a two-dimensional Mari-Kurchan model over a wide range of temperatures. Using the planting technique, we prepared equilibrium configurations down to arbitrarily low temperatures at essentially no cost, well below the MCT crossover temperature, and studied several time correlations functions and their fluctuations.

We have shown that both $\average{C(t)}$ and $\average{F_\text s(t)}$ develop a two-step decay separated by a plateau regime when the glassy dynamics emerges. The final decay occurs even far below the mode-coupling crossover $T_d$, as observed in finite-dimensional glass-formers. The MCT transition is thus strongly avoided in the MK model. Since a Kauzmann transition does not occur either, RFOT theory has little to say about the glassy dynamics in that model, even though it was initially introduced as a glass model with mean-field character. 

Overall, the observed glassy dynamics is quantitatively consistent with the phenomenology of supercooled liquids. The final decay of time correlation functions can be described by a stretched exponential, and a clear excess signal appears at intermediate frequencies which resembles the excess wings found in molecular and simulated liquids. Finally, structural relaxation is accompanied by increasing multi-point dynamic correlations characterising the emergence of collective dynamics at low temperatures. 

Our results show that the MK model displays dynamical properties that are extremely similar to the ones of finite-dimensional systems despite the mean-field nature of its local structure. Because the MK model has no MCT transition, the emergence of collective dynamics cannot be described or attributed to the dynamic criticality. In addition, the thermodynamic Kauzmann transition is forbidden by the random nature of the MK interactions, and it is also impossible to invoke collective activated dynamics over a correlation lengthscale controlled by a decreasing configurational entropy, as envisioned in RFOT theory.   

The sole candidate to explain the emergence of correlated dynamics in the MK model is dynamic facilitation. We expect that the most mobile particles are able to escape their initial position much before the bulk, and this motion then helps neighboring particles (in the shifted representation) to relax more easily themselves. If this process repeats itself, it directly leads to the emergence of dynamic correlations that become very large when structural relaxation finally takes place. In fact, the existence of a small population of particles that relax much faster than the bulk naturally opens a broad time window where dynamic correlations can be built via dynamic facilitation. Moreover, our results suggest that the particles playing the role of the localised defects hypothesized in kinetically constrained models can be statistically identified in the MK model based on their local structure which is, on average, different from the bulk.

The MK model then emerges as an interacting particle model with short-range interactions which, in effect, captures very well the physics of kinetically constrained models. It appears more realistic than the original constrained models themselves which have no Hamiltonian and where defects are introduced by hand. Here, the defects are instead an emergent physical property. The MK model is also different from plaquette models~\cite{garrahan2002glassiness} where interactions do not resemble the ones found in liquids, and the interactions in the MK model are also much less artificial than in models of infinitely-thin hard needles~\cite{Renner1995,Charbonneau2008}.

Since the MK model resembles so closely conventional glass-forming models, it could become a useful tool to better understand how dynamic facilitation can operate in off-lattice models and how to best quantify its consequences. However, it could be that despite the similarities revealed by our work, some more subtle differences emerge after more careful examination. For instance, it is possible that the nature of the particle motion observed at times much shorter than the bulk is simpler in the MK model than in finite-dimensional models where non-trivial thermodynamic fluctuations exist that can influence the slow dynamics. Future efforts should aim at determining whether such differences exist.

In future work, it would also be interesting to test whether the correlation that we have revealed between structure and dynamics is only obeyed at the average level, or whether it applies particle by particle at the microscopic level. Due to the simplified nature of the structure in the MK model, it again appears as a good model to test current efforts aimed at identifying at the microscopic level the correlation between structure and dynamics, which has seen a surge of interest recently due to the advent of machine learning techniques~\cite{Schoenholz2017,Bapst2020,Richard2020}.    

\begin{acknowledgments}
We thank P. Charbonneau and F. Zamponi for discussions. L.B. acknowledges support from the Simons foundation (No. 454933 L. B.). A.I. acknowledges support from JSPS KAKENHI (Grants No. 18H05225, 19H01812, 20H01868, 20H00128). 

\end{acknowledgments}

\bibliography{refs}

\end{document}